# Growth temperature dependence of partially Fe filled MWCNT using chemical vapor deposition

Joydip Sengupta, Chacko Jacob *

Materials Science Centre, Indian Institute of Technology, Kharagpur 721302, India

## ABSTRACT

This investigation deals with the effect of growth temperature on the growth behavior of Fe filled multi-walled carbon nanotubes (MWCNTs). Carbon nanotube (CNT) synthesis was carried out in a thermal chemical vapor deposition (CVD) reactor in the temperature range 650–950 °C using propane as the carbon source, Fe as the catalyst material, and Si as the catalyst support. Atomic force microscopy (AFM) analysis of the catalyst exhibits that at elevated temperature clusters of catalyst coalesce and form macroscopic islands. Field emission scanning electron microscopy (FESEM) results show that with increased growth temperature the average diameter of the nanotubes increases but their density decreases. High-resolution transmission electron microscopy (HRTEM) studies suggest that the nanotubes have multi-walled structure with partial Fe filling for all growth temperatures. The X-ray diffraction (XRD) pattern of the grown materials indicates that they are graphitic in nature. The characterization of nanotubes by Raman spectroscopy reveals that the optimized growth temperature for Fe filled CNTs is 850 °C, in terms of quality. A simple model for the growth of Fe filled carbon nanotubes is proposed.

## 1. Introduction

Research on carbon nanotubes has been extensively carried out in the past few years because of its prospects in basic science and its potential in many technological applications. Owing to their unique mechanical, electrical, thermal, optical, and chemical properties [1], CNTs have become the center of attraction in the field of nanoscale research and are expected to impact key industrial sectors such as nanoelectronics [2], biotechnology [3], and thermal management [4]. Various synthetic methods have been developed for the production of CNTs, including arc-discharge [5], laser ablation [6], pyrolysis [7], and plasma enhanced [8] and thermal [9] CVD. In the last few years, CVD has been the preferred method among different methods because of its potential advantage to produce a large amount of CNTs growing directly on a desired substrate with high purity, large yield, and controlled alignment, whereas the nanotubes must be collected separately in the other growth techniques. Depending on the final application, thermal CVD could be even more desirable than plasma CVD because thermal CVD processes are more economical, suitable for large-area, irregular-shaped substrates, and multiple-substrate coatings [10,11]. Prior to the synthesis of CNTs by CVD, high-temperature hydrogen treatment of the catalyst is an important step in order to produce contamination-free catalyst particles and for the removal of oxides that may exist over the catalyst surface [12].

Another interesting aspect of the CNTs is their cavity, which can be used to incorporate metal clusters in order to generate novel nanostructured materials with new electronic or magnetic properties as a consequence of the large surface-to-volume ratio of the confined materials and the interaction between the confined materials and the inner walls of the nanotube [13,14]. Apart from the geometrical advantage of a cylindrical-shaped nanostructure design, the carbon shells provide an effective protection against the oxidation of the confined metals and ensure their long-term stability at the core. Magnetic metal encapsulated CNTs have extended the potential applications to magnetic force microscopy [15], high-density magnetic recording media [16], and biology [17]. Different filling methods in CNTs including capillary incursion [18], chemical method [19], arc-discharge [20,21], and CVD [22,23] have already been reported.

* Corresponding author. Tel.: +91 3222 283964; fax: +91 3222 255303.
E-mail addresses: joydipdhruba@gmail.com (J. Sengupta), cxj14_holiday@yahoo.com (C. Jacob).

But the processes are costly since they usually require a two-stage system and rigorous control of parameters such as decomposition temperature, heating rate, precursor ratio, etc. The synthesis of filled CNT by a simple one-stage process still remains a major challenge.

Carbon nanotube properties and, consequently, their potential applications strongly depend on the CNTs structural characteristics such as diameter, filling of core or presence of defects. Thus, a very strict control of the experimental parameters is required during the CNT production. But unfortunately, despite tremendous progress in synthesizing CNTs, reports on the systematic comparative study of the growth temperature effect on the filled CNTs are still relatively scarce.

In this paper, we report the results of systematic experiments to study the dependence of nanotube structure on growth temperature for partially Fe filled MWCNTs synthesized from iron using the atmospheric pressure chemical vapor deposition (APCVD) technique. AFM analysis was performed on the Fe films, before and after the heat treatment to observe the resultant change in catalyst surface. The morphology, internal structure and degree of graphitization of carbon nanotubes were investigated using FESEM, XRD, HRTEM, and Raman spectroscopy. To achieve controllable growth of the CNTs, an understanding of their growth mechanism is of importance. Hence, we have also explained the method of partial Fe filling in terms of a growth model that emphasizes the role of the capillary action of the liquid-like iron particles that exist at the time of nanotube nucleation.

## 2. Experimental procedure

Synthesis of Fe filled CNTs was carried out by catalytic decomposition of propane on Si(111) substrates with Fe catalyst in a hot-wall horizontal CVD reactor using a resistance-heated furnace (ELECTROHEAT EN345T). The Si(111) substrates were ultrasonically cleaned with acetone and deionized water prior to catalyst film deposition. A thin film of iron (thickness $\sim 20$ nm) was deposited on the substrate by a vacuum system (Hind Hivac: Model 12A4D) with a base pressure of $10^{-5}$ Torr. The substrates were then loaded into a quartz tube furnace, pumped down to $10^{-2}$ Torr and backfilled with flowing argon to atmospheric pressure. The samples were then heated in argon up to the growth temperature following which the argon was replaced with hydrogen. Subsequently, the samples were annealed in hydrogen atmosphere for 10 min. Finally, the hydrogen was turned off; thereafter propane was introduced into the gas stream at a flow rate of 200 sccm for 1 h for CNT synthesis. The synthesis of Fe filled CNTs was performed at 650, 750, 850, and 950 °C.

An AFM (Nanonics Multiview 1000$^{TM}$) system was used to image the surface morphology of the Fe layer before and after the heat treatment with a quartz optical fiber tip in intermittent contact mode. Samples were also characterized by a Philips X-ray diffractometer (PW1729) with Cu source and $\theta$–$2\theta$ geometry to analyze the crystallinity and phases of grown species. FESEM (ZEISS SUPRA 40) and HRTEM (JEOL JEM 2100) equipped with an energy dispersive X-ray (EDX) analyzer (OXFORD Instruments) were employed for examination of the morphology and microstructure of the CNTs. The sample preparation for high-resolution transmission electron microscopy study was done by scraping the nanotubes from Si substrates, dispersing them ultrasonically in alcohol and then transferring them to carbon-coated copper grids. Raman measurements were carried out at room temperature in a backscattering geometry using a 488 nm air-cooled argon laser as an excitation source for compositional analysis. The Raman spectrometer was equipped with a TRIAX550 single monochromator with a 1200 grooves/mm holographic grating, a holographic super-notch filter and a Peltier-cooled CCD detector.

## 3. Results and discussion

Fig. 1(a) shows the AFM image of the surface profile of the as-deposited Fe film on a Si(111) substrate. The AFM image reveals that the initial film consists of tiny clusters with RMS roughness 3.3 nm. This indicates that the catalyst deposition proceeds via island nucleation and coalescence [24]. Fig. 1(b–e) show the tapping mode AFM images of the surfaces of Fe films after being annealed at 650, 750, 850, and 950 °C in hydrogen. It can be observed that the heat treatment caused the formation of larger islands due to the coalescence of clusters and creates a relatively rough surface. The RMS roughness of the films annealed at 650, 750, 850, and 950 °C are 5.6, 6.7, 9.5, and 10.6 nm, respectively. The coalescence of clusters and formation of macroscopic islands are based on cluster diffusion, which terminates when the island shape is of minimum energy for the specific annealing conditions [25]. The surface diffusion of the clusters increases upon increasing the annealing temperature and in result the sizes of the islands become larger but their density decreases as observed in Fig. 1(b–e) [26,27]. The cluster size plays a critical role in CNT growth.

FESEM was employed for the analysis of the morphology and density of Fe filled CNTs. Fig. 2(a–d) show the FESEM images of carbon nanotubes synthesized at 650, 750, 850, and 950 °C. The CNTs were synthesized on the Si(111) substrates using iron catalyst and the high-aspect ratio nanostructures had randomly oriented spaghetti-like morphology in all the cases. The average diameter of CNTs increases from 25 nm at 650 °C to 77 nm at 950 °C as summarized in Table 1, while the density of the CNTs decreases. With the increase in the growth temperature, the migration rate of iron nanoparticles on the Si surface increases, resulting in significant agglomeration of iron nanoparticles. Consequently due to this agglomeration, at higher temperatures during the thermal CVD process, the size of the iron nanoparticles increases but their density decreases [28], which eventually govern the diameter and density of the grown nanotubes. Fig. 2(e) shows a higher magnification image of the CNTs grown at 750 °C.

For the compositional analysis of the grown species EDX and XRD were performed. Energy dispersive X-ray (Fig. 3(a)) shows that the material contains carbon and iron only, whereas the Si peak is originated from the substrate. In the X-ray diffraction pattern (Fig. 3(b)), the peak at 26.2° is the characteristic graphitic peak arising due to the presence of MWCNTs in the sample. The peak near 43.7° is attributed to the (1 0 1) plane of the nanotube and the peak at 44.7° is from the $\alpha$-Fe phase (JCPDS Card no. 06-0696). The peak at 28.4°, however, is not from the CNTs and is attributed to the (1 1 1) plane of the Si substrate. The XRD pattern confirms the presence of MWCNT, Fe, and Si, which is in agreement with the EDX analysis.

HRTEM was used to characterize the growth morphology and structure dependence of the nanotubes on the growth temperature. Fig. 4(a–d) show the HRTEM images of CNTs synthesized at 650, 750, 850, and 950 °C and illustrate that the nanotubes produced at all the four temperatures are multi-walled. HRTEM studies also reveal that the MWCNTs were produced by the tip growth mechanism, since the catalyst nanoparticles are always encapsulated on the nanotube closed tip (Fig. 4(a–d)). Careful observations show the presence of few elongated nanoparticles encapsulated inside the nanotubes, which demonstrates the hollow nature of the nanotube deposited using iron. Selected area electron diffraction (SAED) pattern of the CNT encapsulated elongated nanoparticle shows spots due to (1 1 0) planes of iron (Fig. 4(e)). The diffractions characteristic to the carbon nanotube (0 0 2) plane originating from the shielding graphitic layers are labeled. Typically, they consist of short arcs, rather than dots, because of imperfectly graphitized tubular walls. The EDX analysis (Fig. 4(f)) of a CNT encapsulated metal nanoparticle shows the presence of Fe and Cu (from the TEM grid), indicating that the encapsulated particle is Fe. There was no evidence for the presence of $Fe_3C$ in the EDX (using TEM) analysis of encapsulated metal particles for all the samples.

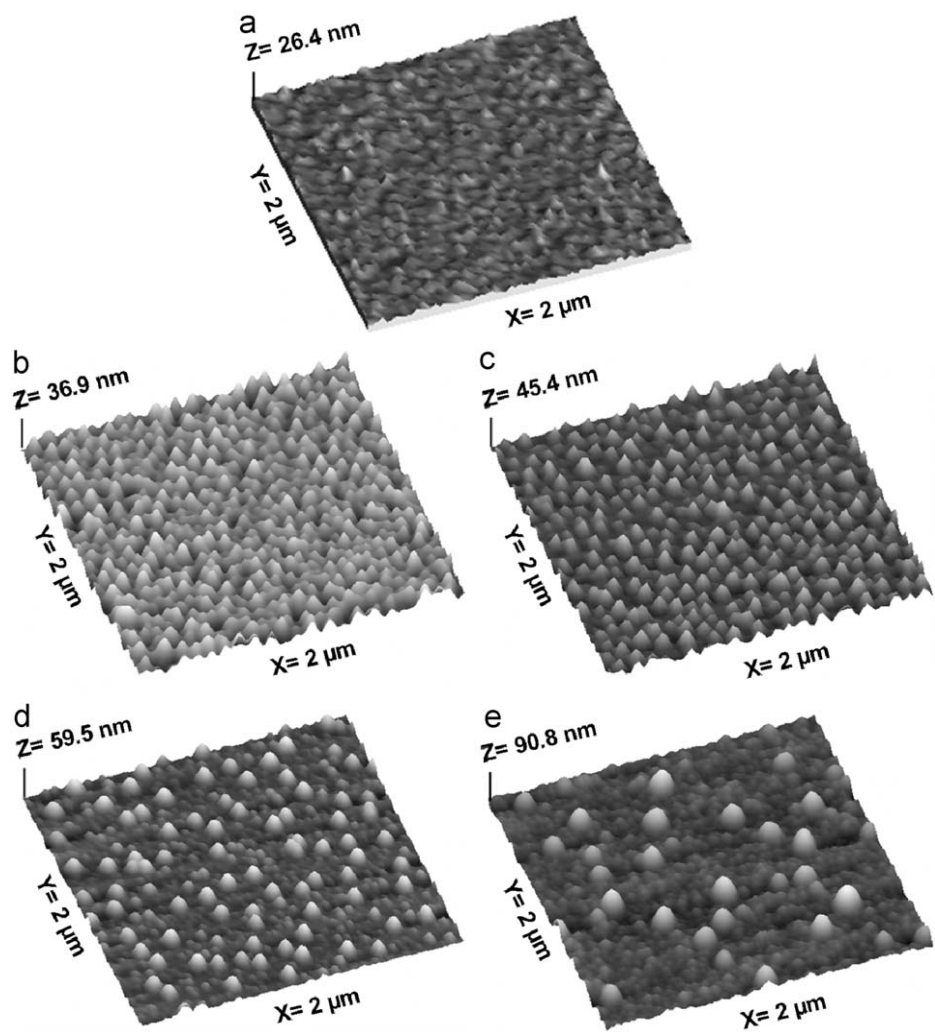

**Fig. 1.** 3-D AFM images of the surfaces of the Fe catalyst on the Si substrates: (a) the as-deposited surface and (b–e) after the thermal treatment at 650, 750, 850, and 950 °C.

To investigate the effect of the growth temperature on the crystallinity of graphite sheets, high-resolution images of CNTs were acquired. The HRTEM image of a carbon nanotube grown at 650 °C (Fig. 5(a)) shows the wavy structure of graphitic sheets at short range. It is well known that the wavy structure is caused due to defects in the graphite sheet. However, the CNT grown at 850 °C (Fig. 5(b)) has clear well-ordered and straight lattice fringes of graphitic sheets separated by 0.34 nm. HRTEM observations indicate that the crystallinity of CNTs improves as the growth temperature increases.

Raman spectroscopy provides more details of the quality and structure of the materials produced. The room-temperature Raman spectra obtained from CNTs that were grown at four different temperatures are shown in Fig. 6. The Raman spectra show two main bands around 1360 cm$^{-1}$ (D band) and around 1585 cm$^{-1}$ (G band), which indicates the presence of MWCNTs. The strong band around 1585 cm$^{-1}$, which is referred as the G band, corresponds to the $E_{2g}$ mode, i.e. the stretching mode of the C–C bond in the graphite plane and demonstrates the presence of crystalline graphitic carbon. The D band at around 1360 cm$^{-1}$ originates from defects in the curved graphitic sheets, tube ends, etc. Beside the D and G band, the spectra of our sample also show another band at ∼1620 cm$^{-1}$ (for 650 and 750 °C), which appears as a shoulder on the G band and is called D′ band. This band accounts for the structural disorder. Like the G band of graphite, the D′ band also corresponds to a graphitic lattice mode with $E_{2g}$ symmetry [29].

Though the D band and G band are present in all CNT samples, there is a difference in peak positions and also a variation in the intensity ratios of these two bands as a function of growth temperature as summarized in Table 2. The observed up-shift of the G band for the nanotubes deposited at 650 and 750 °C temperature can be explained in terms of an overlapping of the G band and D′ band, and a single fit to G+D′ feature gives a net increase of G band position [30]. This suggests that the graphitic carbon films deposited at 650 and 750 °C temperature consist of more disordered structure. The up-shift of the D band in these films can be interpreted as the signature for the increase of disorder in the graphitic structure in accordance with the similar observation from other carbon-based films [31]. It is well known that the $I_D/I_G$ ratio is dependent on the quality of carbon nanotubes and that the D band intensity is also related to the amount of amorphous carbon and defects [32,33]. The highest value of $I_D/I_G$ obtained for CNTs grown at 650 °C indicate that the tubes grown at this temperatures are more defective as compared

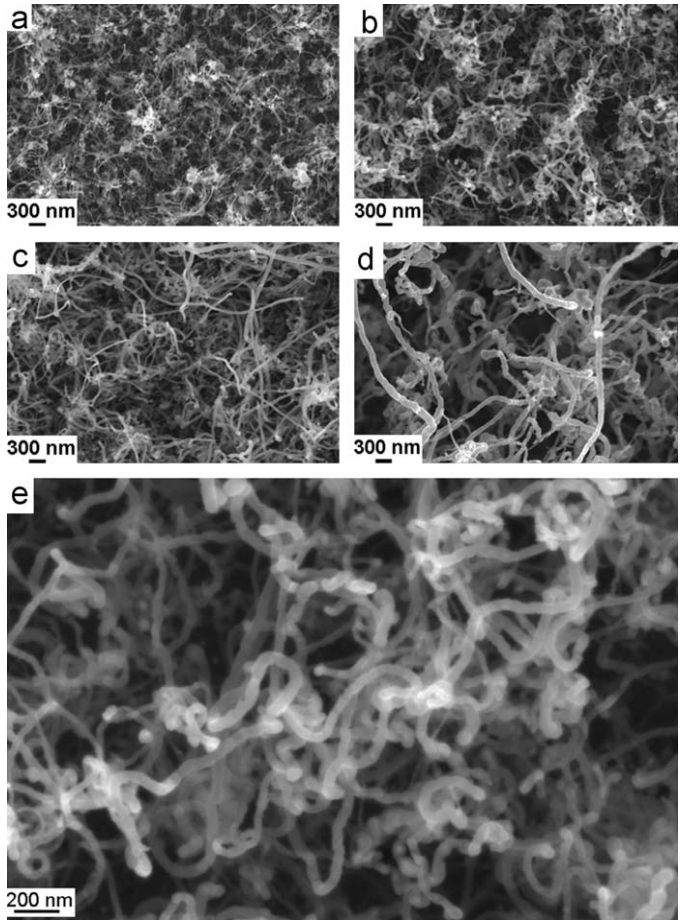

**Fig. 2.** FESEM images of carbon nanotube arrays grown from iron catalyst at different temperatures: (a) 650 °C, (b) 750 °C, (c) 850 °C, and (d) 950 °C; (e) the high magnification FESEM image of the carbon nanotubes grown at 750 °C.

**Table 1**
The diameter distribution and averaged diameter of CNTs grown on iron catalyst using thermal CVD of propane at 650, 750, 850, and 950 °C.

| Temperature (°C) | CNT diameter range (nm) | Average CNT diameter (nm) |
| --- | --- | --- |
| 650 | 10–50 | 25 |
| 750 | 20–70 | 46 |
| 850 | 30–110 | 54 |
| 950 | 30–150 | 77 |

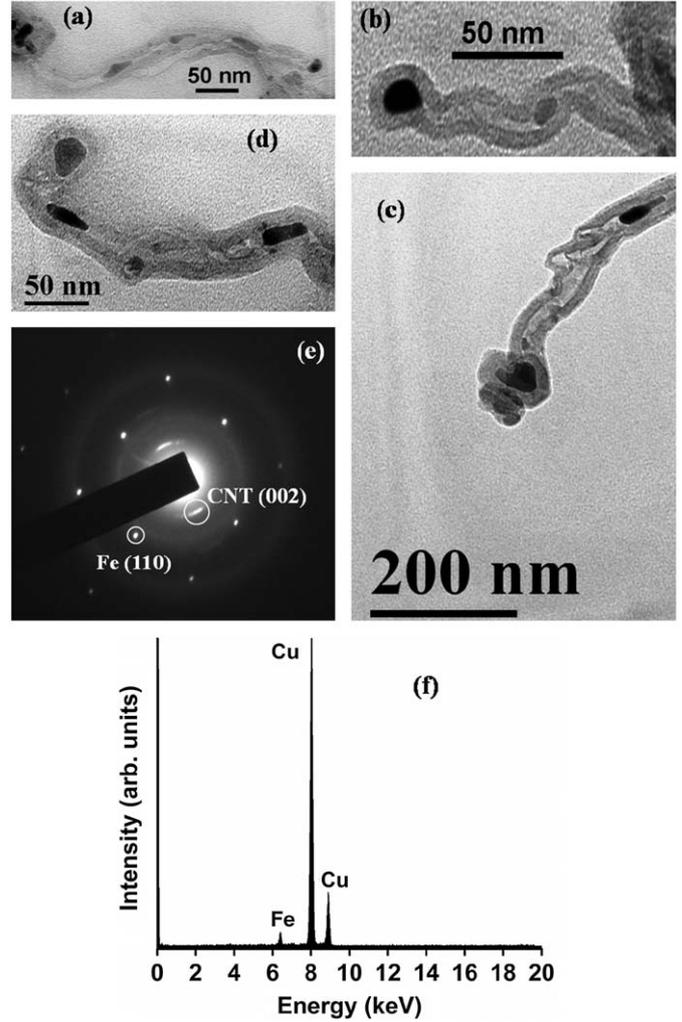

**Fig. 4.** HRTEM images of carbon nanotubes grown from iron catalyst at different temperatures: (a) 650 °C, (b) 750 °C, (c) 850 °C, (d) 950 °C, (e) SAED pattern of a CNT encapsulated elongated iron nanoparticle, and (f) EDX spectrum obtained from a CNT encapsulated metal nanoparticle.

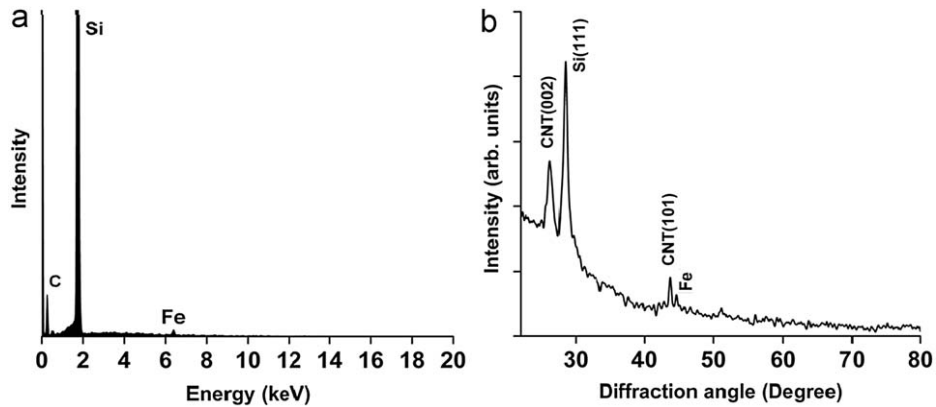

**Fig. 3.** (a) EDX spectrum obtained from the sample grown at 850 °C and (b) XRD spectrum of MWCNTs grown at 850 °C on Si (111) substrate using Fe.

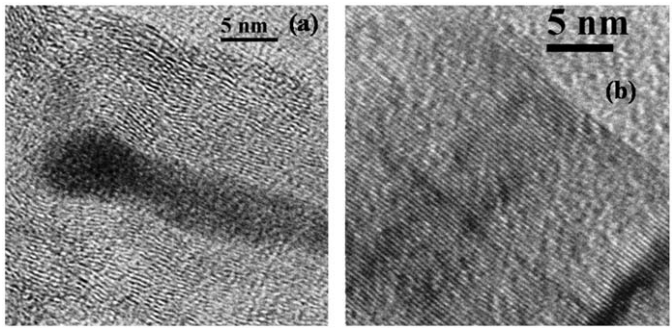

**Fig. 5.** Lattice images from carbon nanotubes grown at (a) 650 °C and (b) 850 °C.

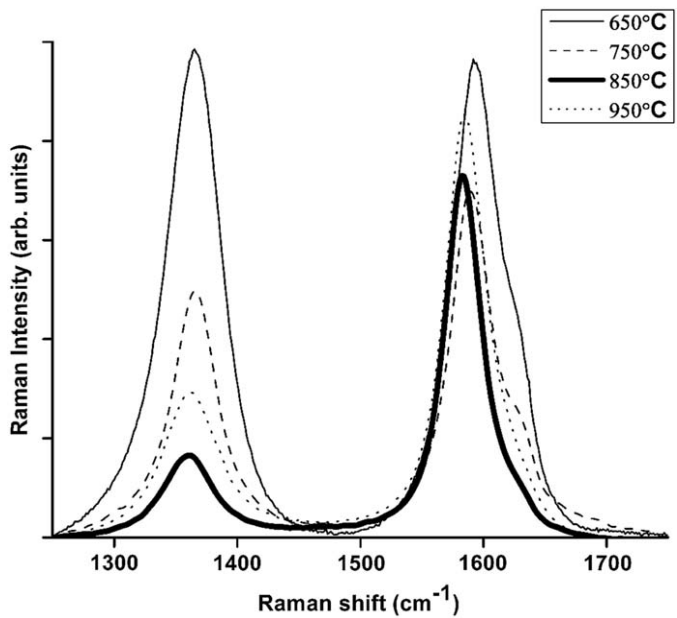

**Fig. 6.** Raman spectra (488 nm excitation) of MWCNTs grown at different temperatures.

**Table 2**
Raman peak positions and D to G band intensity ratios for the CNTs grown at four different temperatures.

| Growth temperature (°C) | D band position (cm$^{-1}$) | G band position (cm$^{-1}$) | $I_D/I_G$ |
|---|---|---|---|
| 650 | 1365.2 | 1591.7 | 1.02 |
| 750 | 1365.1 | 1589.1 | 0.71 |
| 850 | 1361.2 | 1582.7 | 0.23 |
| 950 | 1361.9 | 1584.0 | 0.34 |

to those grown at higher temperatures. Above 650 °C, the relative intensity of the D band to the G band decreases with increasing growth temperature (up to 850 °C), indicating an increasing degree of graphitization in the material. Further increase in the growth temperature results in an increase in the $I_D/I_G$ values, maybe due to the formation of amorphous carbon, which results in a slight increase in the D band, thus resulting in higher $I_D/I_G$ values.

On the basis of the aforementioned experimental results, it can be observed that the as-deposited Fe film contains small clusters indicating Volmer–Weber growth. After high-temperature annealing, the size of the clusters increases but their density decreases. This surface modification occurred due to the increase in the surface diffusion of Fe particles, which depends on growth temperature. With the increase in growth temperature, the surface diffusion of Fe particles over Si substrate increases, resulting in agglomeration of small clusters that eventually produce larger clusters but with a lower density. The density and size of the clusters govern the growth morphology of the CNTs. With the increase in growth temperature the crystallinity of the graphitic sheets also increases and this is responsible for the increase in the degree of graphitization of the CNTs. In all the cases, the catalyst nanoparticles are found both within and at the tip of CNTs, suggesting the tip growth mechanism.

For the synthesis of metal filled CNTs, several authors have used a floating catalyst technique [22,23,34,35], but we have used a fixed catalyst. In our case, the growth model of partially Fe filled CNTs can be explained by the phenomenon of reduction of the melting point of a material with reduction of its size [36–38]. This indicates that the Fe nanoparticles are in a liquid state during the decomposition so that they can easily dissolve the carbon and also diffuse into the nanotube cavities by nanocapillarity [18,39,40], leading to the formation of partially Fe filled CNTs.

## 4. Conclusions

To summarize, we have investigated and optimized the iron catalyzed CVD growth of partially Fe filled CNTs on Si substrates with respect to growth temperature. We find that the average diameter of CNTs increases with growth temperature from 650 to 950 °C, but the density decreases. Moreover, the crystallinity of the CNTs improves progressively with increasing growth temperature up to 850 °C, which is the optimum growth temperature in our case. The results demonstrate that the growth density, diameter and crystallinity of partially Fe filled MWCNTs can be effectively controlled and optimized by adjusting the growth temperature and their growth occurs primarily by the tip growth model based on the capillary action of the liquid-like iron particle.

## Acknowledgements

The authors are grateful to Dr. A. Roy from the Department of Physics and Meteorology, IIT Kharagpur, for her help with the Raman measurement. J. Sengupta is thankful to CSIR for providing the Senior Research Fellowship.

## References

[1] M.S. Dresselhaus, G. Dresselhaus, P. Avouris, Carbon Nanotubes: Synthesis, Structure, Properties and Applications, Springer, Heidelberg, 2001.
[2] P. J-Herrero, J.A.V. Dam, L.P. Kouwenhoven, Quantum supercurrent transistors in carbon nanotubes, Nature 439 (2006) 953–956.
[3] C.R. Martin, P. Kohli, The emerging field of nanotube biotechnology, Nat. Rev. Drug Discovery 2 (2003) 29–37.
[4] W. Kim, R. Wang, A. Majumdar, Nanostructuring expands thermal limits, Nanotoday 2 (2007) 40–47.
[5] D.S. Bethune, C.H. Kiang, M.S. de Vries, G. Gorman, R. Savoy, J. Vazquez, R. Beyers, Cobalt-catalysed growth of carbon nanotubes with single-atomic-layer walls, Nature 363 (1993) 605–607.
[6] W.K. Maser, E. Munoz, A.M. Benito, M.T. Martinez, G.F. de la Fuente, Y. Maniette, E. Anglaret, J.-L. Sauvajol, Production of high-density single-walled nanotube material by a simple laser-ablation method, Chem. Phys. Lett. 292 (1998) 587–593.
[7] M. Terrones, N. Grobert, J. Olivares, J.P. Zhang, H. Terrones, K. Kordatos, W.K. Hsu, J.P. Hare, P.D. Townsend, K. Prassides, A.K. Cheetham, H.W. Kroto, D.R.M. Walton, Controlled production of aligned-nanotube bundles, Nature 388 (1997) 52–55.
[8] Z.F. Ren, Z.P. Huang, J.W. Xu, J.H. Wang, P. Bush, M.P. Siegel, P.N. Provencio, Synthesis of large arrays of well-aligned carbon nanotubes on glass, Science 282 (1998) 1105–1107.
[9] S. Fan, M.G. Chapline, N.R. Franklin, T.W. Tombler, A.M. Cassell, H. Dai, Self-oriented regular arrays of carbon nanotubes and their field emission properties, Science 283 (1999) 512–514.

[10] Y.C. Choi, D.W. Kim, T.J. Lee, C.J. Lee, Y.H. Lee, Growth mechanism of vertically aligned carbon nanotubes on silicon substrates, Synth. Met. 117 (2001) 81–86.
[11] A. Huczko, Synthesis of aligned carbon nanotubes, Appl. Phys. A 74 (2002) 617–638.
[12] D. Takagi, H. Hibino, S. Suzuki, Y. Kobayashi, Y. Homma, Carbon nanotube growth from semiconductor nanoparticles, Nano Lett 7 (2007) 2272–2275.
[13] S. Karmakar, S.M. Sharma, M.D. Mukadam, S.M. Yusuf, A.K. Sood, Magnetic behavior of iron-filled multiwalled carbon nanotubes, J. Appl. Phys. 97 (2005) 054306(1–5).
[14] J. Sloan, A.I. Kirkland, J.L. Hutchison, M.L.H. Green, Integral atomic layer architectures of 1D crystals inserted into single walled carbon nanotubes, Chem. Commun. 13 (2002) 1319–1332.
[15] A. Winkler, T. Mühl, S. Menzel, R.K. Koseva, S. Hampel, A. Leonhardt, B. Büchner, Magnetic force microscopy sensors using iron-filled carbon nanotubes, J. Appl. Phys. 99 (2006) 104905(1–5).
[16] C.T. Kuo, C.H. Lin, A.Y. Lo, Feasibility studies of magnetic particle-embedded carbon nanotubes for perpendicular recording media, Diamond Relat. Mater. 12 (2003) 799–805.
[17] E.B. Palen, Iron filled carbon nanotubes for bio-applications, Mater. Sci.—Poland 26 (2008) 413–418.
[18] P.M. Ajayan, S. Iijima, Capillarity-induced filling of carbon nanotubes, Nature 361 (1993) 333–334.
[19] S.C. Tsang, Y.K. Chen, P.J.F. Harris, M.L.H. Green, A simple chemical method of opening and filling of carbon nanotubes, Nature 372 (1994) 159–162.
[20] A. Loiseau, H. Pascard, Synthesis of long carbon nanotubes filled with Se, S, Sb and Ge by the arc method, Chem. Phys. Lett. 256 (1996) 246–252.
[21] C. Guerret-Plecourt, Y.L. Bouar, A. Loiseau, H. Pascard, Relation between metal electronic structure and morphology of the compounds inside carbon nanotubes, Nature 372 (1994) 761–765.
[22] A. Leonhardt, M. Ritschel, R. Kozhuharova, A. Graff, T. Muhl, R. Huhle, I. Monch, D. Elefant, C.M. Schneider, Synthesis and properties of filled carbon nanotubes, Diamond Relat. Mater. 12 (2003) 790–793.
[23] C. Müller, D. Golberg, A. Leonhardt, S. Hampel, B. Büchner, Growth studies, TEM and XRD investigations of iron-filled carbon nanotubes, Phys. Status Solidii (A) 203 (2006) 1064–1068.
[24] W.A. Tiller, The Science of Crystallization: Microscopic Interfacial Phenomena, Cambridge University Press, Cambridge, 1991.
[25] M.J.J. Jak, C. Konstapel, A. van Kreuningen, J. Verhoeven, J.W.M. Frenken, Scanning tunnelling microscopy study of the growth of small palladium particles on TiO$_2$(110), Surf. Sci. 457 (2000) 295–310.
[26] S. Pisana, M. Cantoro, A. Parvez, S. Hofmann, A.C. Ferrari, J. Robertson, The role of precursor gases on the surface restructuring of catalyst films during carbon nanotube growth, Physica E 37 (2007) 1–5.
[27] C. Muller, A. Leonhardt, M.C. Kutz, B. Büchner, H. Reuther, Growth aspects of iron-filled carbon nanotubes obtained by catalytic chemical vapor deposition of ferrocene, J. Phys. Chem. C 113 (2009) 2736–2740.
[28] R.M. Yadav, P.S. Dobal, T. Shripathi, R.S. Katiyar, O.N. Srivastava, Effect of growth temperature on bamboo-shaped carbon–nitrogen (C–N) nanotubes synthesized using ferrocene acetonitrile precursor, Nanoscale Res. Lett. 4 (2009) 197–203.
[29] A. Sadezky, H. Muckenhuber, H. Grothe, R. Niessner, U. Poschl, Raman microspectroscopy of soot and related carbonaceous materials: spectral analysis and structural information, Carbon 43 (2005) 1731–1742.
[30] A.C. Ferrari, J. Robertson, Interpretation of Raman spectra of disordered and amorphous carbon, Phys. Rev. B 61 (2000) 14(095–107).
[31] S. Choi, K.H. Park, S. Lee, K.H. Koh, Raman spectra of nano-structured carbon films synthesized using ammonia-containing feed gas, J. Appl. Phys. 92 (2002) 4007–4011.
[32] F. Tunistra, J.L. Koenig, Raman spectrum of graphite, J. Chem. Phys. 53 (1970) 1126–1130.
[33] R. Saito, G. Dresselhaus, M.S. Dresselhaus, Physical Properties of Carbon Nanotubes, Imperial College Press, London, 1998.
[34] F. Geng, H. Cong, Fe-filled carbon nanotube array with high coercivity, Physica B 382 (2006) 300–304.
[35] X. Zhang, A. Cao, B. Wei, Y. Li, J. Wei, C. Xu, D. Wu, Rapid growth of well-aligned carbon nanotube arrays, Chem. Phys. Lett. 362 (2002) 285–290.
[36] P.-A. Baffat, Lowering of the melting temperature of small gold crystals between 150 Å and 25 Å diameter, Thin Solid Films 32 (1976) 283–286.
[37] C. Pan, Z. Zhang, X. Su, Y. Zhao, J. Liu, Characterization of Fe nanorods grown directly from submicron-sized iron grains by thermal evaporation, Phys. Rev. B 70 (2004) 233404(1–4).
[38] J.M. Thomas, P.L. Walker, Mobility of metal particles on a graphite substrate, J. Chem. Phys. 41 (1964) 587–588.
[39] J. Fujita, M. Ishida, T. Ichihashi, Y. Ochiai, T. Kaito, S. Matsui, Carbon nanopillar laterally grown with electron beam-induced chemical vapor deposition, J. Vac. Sci. Technol. B 21 (2003) 2990–2993.
[40] T. Ichihashi, J. Fujita, M. Ishida, Y. Ochiai, In situ observation of carbon-nanopillar tubulization caused by liquid-like iron particles, Phys. Rev. Lett. 92 (2004) 215702 (1–4).